# Eye Tracking for Tele-robotic Surgery: A Comparative Evaluation of Head-worn Solutions

Regine Büter*[a], Roger D. Soberanis-Mukul[a], Paola Ruiz Puentes[a], Ahmed Ghazi[b], Jie Ying Wu[c], Mathias Unberath[a]

[a]Department of Computer Science, Johns Hopkins University, 3400 N Charles St, Baltimore, Maryland 21218, US; [b]Department of Urology, Johns Hopkins Medical Institute, 600 N Wolfe St, Baltimore, Maryland 21287, USA; [c]Department of Computer Science, Vanderbilt University, 2201 West End Ave, Nashville, Tennessee 37235, USA

## ABSTRACT

**Purpose:** Metrics derived from eye-gaze-tracking and pupillometry show promise for cognitive load assessment, potentially enhancing training and patient safety through user-specific feedback in tele-robotic surgery. However, current eye-tracking solutions' effectiveness in tele-robotic surgery is uncertain compared to everyday situations due to close-range interactions causing extreme pupil angles and occlusions. To assess the effectiveness of modern eye-gaze-tracking solutions in tele-robotic surgery, we compare the Tobii Pro 3 Glasses and Pupil Labs Core, evaluating their pupil diameter and gaze stability when integrated with the da Vinci Research Kit (dVRK). **Methods:** The study protocol includes a nine-point gaze calibration followed by pick-and-place task using the dVRK and is repeated three times. After a final calibration, users view a 3x3 grid of AprilTags, focusing on each marker for 10 seconds, to evaluate gaze stability across dVRK-screen positions with the L2-norm. Different gaze calibrations assess calibration's temporal deterioration due to head movements. Pupil diameter stability is evaluated using the FFT from the pupil diameter during the pick-and-place tasks. Users perform this routine with both head-worn eye-tracking systems. **Results:** Data collected from ten users indicate comparable pupil diameter stability. FFTs of pupil diameters show similar amplitudes in high-frequency components. Tobii Glasses show more temporal gaze stability compared to Pupil Labs, though both eye trackers yield a similar 4cm error in gaze estimation without an outdated calibration. **Conclusion:** Both eye trackers demonstrate similar stability of the pupil diameter and gaze, when the calibration is not outdated, indicating comparable eye-tracking and pupillometry performance in tele-robotic surgery settings.

## 1. INTRODUCTION

Gaze estimation and pupillometry are increasingly important areas in computer-assisted interventions, particularly in the context of tele-robotic surgery. By analyzing various aspects derived from eye-related metrics such as gaze, pupil diameter, and blink rate, it has been suggested that cognitive load of users may be effectively assessed[1-3]. Understanding the user's cognitive effort during tele-robotic surgery provides valuable information regarding the user's skill level [4]. This knowledge can be used to define tailored training programs for trainees, ultimately leading to improved learning experiences. In turn, patient safety can be enhanced by reducing the mental workload of surgeons, increasing their ability to handle unexpected situations effectively [5].

Different eye-tracking systems have been used to estimate cognitive load for various scenarios. For example, in their studies, Wu et al. utilized the Tobii Glasses Pro 2 (Tobii, Stockholm, Sweden) to assess workload during robotic surgical skill training. They employed gaze entropy and pupil diameter as indicators of workload in robotic skill simulation sessions [4]. In contrast, Melnyk et al. used an eye tracker from Pupil Labs (Berlin, Germany) to implement gaze augmented training in robotic surgery by using expert gaze patterns to improve trainee performance and efficiency [6].

Challenges emerge when using eye-tracking solutions in tele-robotic surgery environments compared to everyday scenarios. The main challenges for eye-tracking are the extreme downward angle of the pupil and occlusion. These challenges result from the fact that surgical maneuvers are performed at close working distance requiring large eye movements to traverse the workspace, and the configuration of the surgeons position in front of the console. In the da Vinci Surgical System configuration, the surgeon's head is supported by the console at an upright to slightly downward angle, causing them to have a downward view on the stereo vision video.

Because the performance and usability of any eye data-based cognitive effort detection algorithm directly depends on the quality of the eye and pupil signals, ensuring adequate data quality is the first step in enabling this technology. Thus, the main goal of this study is to compare the eye-tracking and pupillometry capabilities of the Tobii Pro 3 Glasses and the

Pupil Labs eye-tracking solution in the da Vinci Research Kit (dVRK) with a matched-user study. Specifically, we assess the stability of the gaze tracking and the pupil diameter estimation, both of which are important indicators of cognitive load. This analysis aims to provide insights into the strengths and limitations of both systems for eye-tracking and pupillometry applications.

## 2. SPECIFICATION OF THE EYE TRACKERS

This study compares two head-worn eye-tracking solutions, namely the Tobii Pro Glasses 3 and the Pupil Labs Core.
A console-mounted setup was also considered in this study, however head-worn solutions are preferred due to some limitations. These limitations include facing difficulties in achieving optimal camera angles for accurate pupil tracking and the hindrance of the pupil detection by the presence of eyelashes, which is less problematic in head-mounted solutions. Furthermore, console-mounted setups impose restrictions on head movement due to the calibration of the eye-balls, because the head-mounted solutions maintain a relatively constant eye-to-camera distance, assuming minimal glasses slippage. Due to these difficulties, we perform our analysis on the head-worn solutions.

The Tobii Pro Glasses 3, manufactured by Tobii (Stockholm, Sweden), has two cameras per eye with a sampling rate of 100Hz. The glasses also include an inertial measurement unit (IMU) and a 16-bit mono microphone. Gaze information is obtained with a one-point calibration procedure performed before the recording begins. Recorded data includes 2D and 3D gaze points, pupil diameter, and 3D gaze origin for both eyes. Scene camera view and microphone audio are also recorded if enabled. Metadata is also available, e.g. the creation time of the recording in UTC or the calibration information of the scene camera. The Tobii Labs Pro software is only available for Windows [7], whereas the Software Tobii Pro Glasses Controller is available for Windows, Mac, and Android, with an API for developers that provides a web socket with functionality the control the glasses [8]. A Tobii Pro SDK for the Glasses is available as the Python library g3pylib, which packages the API for easier use in custom applications, though the Tobii Software is not open-source [9].
The Pupil Labs Core, produced by Pupil Labs (Berlin, Germany), uses one camera per eye at a 200Hz sampling frequency. There are no additional built-in sensors. Calibration can be done primarily using a five-marker pattern, or alternative methods like user-defined landmarks. The calibration can be performed before or after the recording using the collected data (Post-Hoc Calibration) [10]. Recorded data includes eye video files, gaze data, pupil diameter for both eyes, and world video data. The eye tracker's software, Pupil Capture, Pupil Player, and Pupil Service, is open-source and available for Windows, Linux, and MacOS. An API is provided for control via a web socket in custom applications [11].

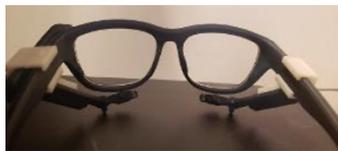

Figure 1. The Pupil Core cameras mounted to the Pupil Invisible Glasses.

For integration into the dVRK-Console, we mounted the Pupil Core cameras on the Pupil Labs Invisible glasses with customized 3D-printed pieces, as seen in Figure 1. Through the customizability of the cameras due to the 3D printed piece, we can ensure that the eyes are in good view of the cameras.

## 3. METHODS

A matched-user study compared the stability of the gaze and pupil diameter of the eye-tracking solutions in the dVRK. Ten participants were recruited at Johns Hopkins University (approved under HIRB00014648), to conduct a user study with the Tobii Glasses and the Pupil Labs eye tracker. The following study procedure was conducted:
Calibration of the Tobii Glasses is accomplished through the one-point calibration procedure integrated into the Tobii Pro Controller Software. It is important to note that this calibration is exclusive to the Tobii Glasses and not required for the Pupil Labs eye tracker. In the experiment procedure, valid for both eye trackers, first, a calibration is performed, followed by a pick and place task (Peg-Transfer). This process is repeated three times, followed by a fourth calibration and the use of a grid of AprilTags to estimate the calibration error.
The calibration procedure involves a 9-point stereo calibration video displayed on the dVRK console, with each marker shown for 5 seconds, with start and end screens. The video lasts 55 seconds. The recorded data is used to map the gaze from the Tobii Glasses to the dVRK screen via rigid registration [12] and scaling, as there is no calibration possible after the recording. On the other hand, the Pupil Labs eye tracker is calibrated to dVRK with the same 9-point SPAAM procedure using Pupil Player software. As both eye trackers handle monocular videos, the gaze is mapped to the right or left video stream. To evaluate the calibration, a 3x3 grid of AprilTags is shown through the endoscope camera. The user is asked to focus on each tag for 10 sec. To estimate the calibration error, the L2-norm between the gaze point and the AprilTag is calculated. Angular accuracy couldn't be assessed due to unknown user-screen distance from dVRK console's structure. The comparison of the stability of the detected pupil diameter is based on the Fourier transform of the data collected during the Peg-Transfer task.

As the saved UTC-creation time of the Tobii Glasses does not map to the system time of the computer used for the dVRK, temporal synchronization was done manually by estimating the time difference in seconds. Therefore, the temporal unsynchronized gaze, that should be collected during the gaze evaluation step was viewed as a video. From this video it can be observed when the gaze should switch to the next AprilTag on the grid. The time difference is the time from when we know the change should happen and when it can be observed it changes in the video.

## 4. PRELIMINARY RESULTS

To evaluate the stability of the measurement of the pupil diameter, the Fast Fourier Transform (FFT) is estimated from the measured pupil diameter during the Peg-Transfer Tasks. In the high-frequency components of the FFT, we can observe the stability of the pupil diameter estimation for each eye tracker. If one eye tracker's detection is less stable than another's, the amplitude in these high frequencies will be lower for the less stable eye tracker. The results (comp. Figure 5) with regards to the Tobii Glasses show that the amplitude of the right FFT is overall lower than the one from the left eye. For the Pupil Labs (comp. Figure 4) eye tracker both FFTs are nearly identical. Overall, the FFT show similar results in both trackers, with similar amplitudes of the FFT in the high and low frequencies.

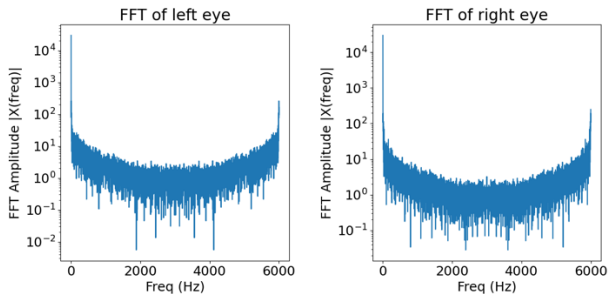

Figure 5. The FFT from the pupil diameter of the Tobii Pro 3 Glasses of the left and right diameter over all users with a log scale.

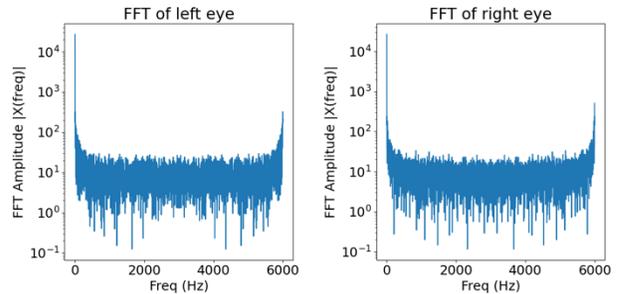

Figure 4. The FFT from the pupil diameter of the Pupil Labs eye tracker of the left and right diameter over all users with a log scale.

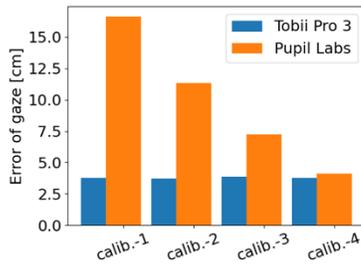

Figure 2. The error of the gaze (cm) on each the AprilTag Grid with different mapping/calibrations for the Tobii Glasses and the Pupil Labs eye tracker.

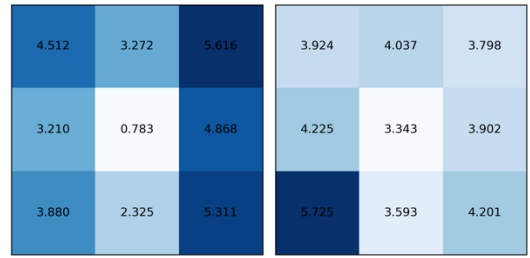

Figure 3. Heatmap of mean errors (cm) over all users for different areas of the surgeon console display. Right of the Tobii Glasses and left is from the Pupil Labs eye tracker.

In order to evaluate the effectiveness of the eye tracker with regards to the gaze mapped to the dVRK video, we estimate the error of the calibrated or mapped gaze point to the respective AprilTag. Different calibration steps are used, to analyze the deterioration of the calibration over time, as the calibrations took place at different places in time each followed by a Peg-Transfer task to mimic the use of the dVRK. In Figure 2 it can be seen, that the Tobii Glasses have approximately the same error for all calibration steps, whereas the Pupil Labs eye tracker has a high error when the first calibration is used to calibrate the gaze, with the error decreasing with each step, to approximately the same error as the Tobii Glasses. For the last calibration step, both have an error of around 4cm.

Figure 3 shows the gaze tracking systems' performance across screen sections. For the Tobii Glasses it comes apparent, that the right third of the screen cannot be detected as accurate as the middle or left hand of the screen. The Tobii eye tracker detects the gaze with a low error, when the user looks in the middle of the screen. For the Pupil Labs eye tracker, the highest error is in the bottom row of the grid.

## 5. CONCLUSION

The results of the Fast Fourier Transform (FFT) analysis for both the Tobii Glasses and the Pupil Labs eye tracker indicate that their performance in detecting the pupil diameter is similar, as the stability of gaze tracking the high-frequency components of the FFTs are comparable. Although the mid frequencies are slightly lower in the Tobii Glasses, the amplitudes are generally similar in both devices.

Results of the gaze analysis show, that without an outdated calibration of the Pupil Labs eye tracker, the error of the gaze estimation is comparable to the estimation of the Tobii Glasses. The error distribution of the error over the screen is more evenly with the Pupil Labs eye tracker, while the Tobii Glasses exhibit lower error in the middle but higher error at the edges. This unequal distribution of the error of the gaze might be due to the extreme angles the user's pupil encounters during the use of the dVRK. As this unequal distribution could not be observed in the pupil labs eye tracker with the custom 3D piece, the customizability of the eye tracker can encounter the extreme angles of the users pupils insight the dVRK. Additionally, the Tobii Glasses require a workaround to map the gaze onto the dVRK video, unlike the direct calibration capability of the Pupil Labs solution.

Overall, the Tobii Pro 3 Glasses and the Pupil Labs eye tracker exhibit similar results for the pupil diameter estimation and the gaze estimation after a recent calibration. Though the Tobii Glasses have a more stable gaze estimation during the use of the dVRK, the disadvantage of the Tobii Glasses is its proprietary solution, which prevents a direct calibration of the gaze to the dVRK video.

## 6. NEW OR BREAKTHROUGH WORK TO BE PRESENTED

Our study demonstrates that both the Tobii Pro 3 Glasses and the Pupil Labs eye tracker exhibit similar performance in estimating pupil diameter and gaze during tele-robotic surgery. Moreover, we reveal that both eye trackers encounter comparable challenges in accurately estimating the gaze point, which is influenced by the user's positioning within the console.

## ACKNOWLEDGMENTS

We would like to thank Tobii for making available Tobii Pro 3 glasses used for data collection. This work was funded in part by an Intuitive Surgical Technology Research Grant and Johns Hopkins University internal funds.